# Structural dimerization and charge-orbital ordering in ferromagnetic semiconductor LiV$_2$S$_4$ monolayer

Rui Song,[a] Bili Wang,[a] Kai Feng,[a] Jia Yao,[a] Mengjie Lu,[a] Jing Bai,[a] Shuai Dong[b] and Ming An*[b]

With the rise of two-dimensional (2D) materials, unique properties that are completely distinct from bulk counterparts continue to emerge at low-dimensional scales, presenting numerous opportunities and challenges. It also provides a new perspective for the study of transition metal system. Here, based on density functional theory (DFT), the physical properties of 2D monolayer LiV$_2$S$_4$ have been studied. Remarkable changes have been observed, i.e., vanadium dimerization, ferromagnetism, charge distribution and metal-insulator transition (MIT). It is argued that the electronic instability leads to the V dimerization which further lifts the degeneracy of charge distribution and stabilizes the charge and spin ordering state.

## Introduction

Transition metal compounds with multi-degrees of freedom (charge, spin, orbital and lattice) often emerge abundant properties such as charge/orbital ordering, MIT, electronic phase separation, multiferroicity, colossal magnetoresistance, high-temperature superconductivity.[1-4] These phenomena and their underlying mechanisms are important issues involving many fields of condensed-matter physics, which have been extensively studied for more than a century. Nowadays, with the rise of two-dimensional (2D) materials,[5] novel characteristics that are absent in the bulk form have been found and various breakthroughs have been witnessed, ushering in a new era of low-dimensional materials.[6-9] Apparently, low dimensionality also provides an ideal platform for the study of transition metal system, since it can lead to quantized electron states and thus offers new approaches for property regulation.[10-16] Actually, the intrinsic magnetism and ferroelectricity discovered in the currently known atomic-thin transition metal materials have already shown clear evidence for the enormous scientific value and application potential of low-dimensional transition metal system.[7,17]

However, in contrast to the high enthusiasm for 2D materials, the research on low-dimensional transition metal compounds are relatively limited, which mainly comes from the fact that most transition metal compounds exist in three-dimensional bulk form, rather than the graphene-like, i.e., layered van der Waals (vdW) stacking, structure. Fortunately, due primarily to technological advances, more and more 2D materials have been prepared, including non-vdW transition metal compounds.[18-21] These nanosheets can be obtained by mechanical peeling,[22] selective etching[23] or ion intercalation[19,24] with their thickness down to monolayer limit, making the study of non-vdW transition metal compounds possible.[25]

Recently, AgCr$_2$S$_4$ single layer has been successfully exfoliated from its non-vdW parent bulk,[19] with its composition and crystal framework well preserved. This successful peeling greatly inspires further research on the low-dimensional properties of non-vdW materials, especially those belonging to the AMX$_2$ family (where A is a monovalent metal, M is a transition metal and X is a chalcogen).[20,26] AMX$_2$ compounds have a layered structure composed of a stacking of spacer A layers and MX$_2$ layers. The latter is composed of edge-sharing MX$_6$ octahedra forming geometrically frustrated triangular lattice. Both the specific stacking pattern and the A-site coordination can lead to distinct structural symmetries.[27,28] In addition to the structural variety, the multiple electronic degrees of freedom often give rise to intriguing behaviors such as cluster formation,[29,30] spin/orbital ordering[2,27,31] and superconductivity.[32]

Here, based on first-principles calculations, the basic properties of bulk LiVS$_2$ and the novel behavior exhibited by single layer LiV$_2$S$_4$ with dimensionality reduction have been investigated. Bulk LiVS$_2$ crystallizes in a trigonal structure, its layered feature is illustrated in Fig. 1a. Both Li and V ions are located in the octahedral holes of hexagonal close-packed S atoms, forming 2D triangular lattices (see Fig. 1b). Upon cooling, a paramagnetic metal to nonmagnetic insulator transition takes place at about 310 K,[30,33] accompanied by in-plane vanadium trimerization. Moreover, being sandwiched between insulating LiVO$_2$ and metallic LiVSe$_2$, LiVS$_2$ is believed to locate at the boundary between correlated insulator and metal.[29] Based on the above facts, low-dimensional effects may provide an additional parameter for manipulating the underlying interactions and therefore may possibly lead to unique collective behaviors.

[a.] Department of General Education, Army Engineering University of People's Liberation Army, Nanjing 211101, China.
[b.] School of Physics, Southeast University, Nanjing 211189, China. E-mail: amorn@seu.edu.cn

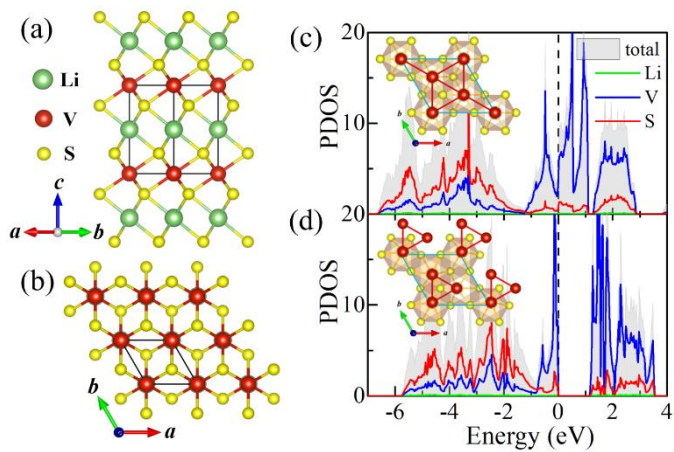

Fig.1 (a) Side and (b) top views of bulk LiVS$_2$, the unit cell is denoted with black lines. The Li, V and S atoms are depicted by the green, red and yellow balls, respectively. (c) The atomic projected density of states (PDOS) of the highly symmetric LiVS$_2$ with ideal triangular lattice (P-3m1 phase). (d) The PDOS of LiVS$_2$ with trimerized lattice (P31m phase). The corresponding structures are illustrated in the insets.

In this work, the structural, magnetic, and electronic properties of LiV$_2$S$_4$ monolayer have been studied. Drastic changes have been observed, i.e., appearance of magnetism, charge ordering, vanadium dimerization along with MIT. At single layer limit, novel mixed-valent state emerges with nominally 1:1 ratio of V$^{3+}$ (3$d^2$) and V$^{4+}$ (3$d^1$). We suggest that the partially occupied $d$ orbitals and the coupling with electronic degrees of freedom are responsible for lifting the degeneracy of charge distribution and stabilizing the charge ordering state.

## Methods

Our DFT calculations were performed using Vienna ab initio Simulation Package known as VASP.[34,35] The electron-ion interaction was described by projected augmented-wave (PAW) pseudo-potentials, with semicore states treated as valence states.[36] The exchange and correlation were treated using Perdew-Burke-Ernzerhof (PBE) parametrization of the generalized gradient approximation (GGA).[37] To properly describe the correlated electrons, the GGA+$U$ method was adopted, and the on-site Hubbard $U_{eff}$ was imposed on V's 3$d$ orbitals using the Dudarev approach for all calculations. The plane-wave cutoff energy was set to 650 eV. The Monkhorst-Pack scheme is chosen to sample the Brillouin zone, with 9×9×5 $k$ point for the primitive cell of bulk LiVS$_2$, 5×5×1 and 4×5×1 for monolayer LiV$_2$S$_4$ with rhombohedral and orthorhombic superstructure, respectively. For monolayers, a 20 Å vacuum layer was added along the $c$-direction to avoid the interaction between adjacent slices. The energy convergent criterion was set to 1×10$^{-5}$ eV. Both the in-plane lattice parameters and the internal atomic coordinates were fully optimized with the conjugate gradient algorithm implemented in VASP until the residual Hellman-Feynman forces were less than 0.01 eV/Å. The phonon band structure was calculated using the density functional perturbation theory (DFPT).[38,39]

## Results and discussion

Firstly, theoretical studies on parent bulk have been carried out. According to previous report on bulk LiVS$_2$,[30,33] the high temperature phase belongs to P-3m1 space group with equispaced V ions forming perfect triangular lattice. At around 310 K, vanadium clusters were observed with a trimer configuration exhibiting P31m symmetry. Based on our first-principles calculations, this structural phase transition has been confirmed by taking the ideal triangular lattice (P-3m1 phase) and the trimerized lattice (P31m phase) into consideration. By comparing the total energies at 0 K, the P31m phase is energetically more favorable, whose DFT energy is 331 meV/V lower than that of the undistorted P-3m1 phase, in agreement with previous experimental observations.[30,33]

As shown in Fig. 1c and d, the accompanying MIT has also been verified by our calculations. The top view of corresponding structures is illustrated in the insets. It can be seen that the in-plane vanadium trimerization not only lowers LiVS$_2$'s structural symmetry, from P-3m1 (space group No. 164) to P31m (No. 157), but also opens an energy gap (1.15 eV). This MIT accompanied by in-plane trimerization is consistent with previous observations reported in refs. 29,33. The electronic densities of states (DOS) near the Fermi level mainly originate from V's 3$d$ orbitals, exhibiting Mott-Hubbard insulating behavior.

The influence of the effective Hubbard $U$ ($U_{eff}$) has also been considered (see Fig. S1 in the supplementary material). The in-plane (out of plane) lattice constant increases (decreases) monotonically with the increase of $U_{eff}$. The optimized lattice constants are in good agreement with the previously reported values when $U_{eff}$ is taken as 3.5 eV.[27,30] Additionally, the influence of $U_{eff}$ on the electroinc properties of both LiVS$_2$ bulk and LiV$_2$S$_4$ monolayer has been investigated with the commonly adopted value ranging from 2 to 4 eV. The obtained results have no substantial impact on our final conclusion. Therefore, this specific $U_{eff}$ value (3.5 eV) will be adopted in the following calculations by default.

Inspired by the successful exfoliation of AgCr$_2$S$_4$ monolayer via ion intercalation,[19] LiVS$_2$ with similar layered structure may likewise be exfoliated into nanosheets and even monolayers. As mentioned above, dimensional reduction is expected to lead to unique properties and help reveal the underlying mechanisms behind these phenomena. Therefore, the structure, magnetic and electronic properties, and the stability of LiV$_2$S$_4$ monolayer have been explored in detail in the following.

Starting from the highly symmetric P-3m1 bulk phase, a LiV$_2$S$_4$ monolayer has been constructed. As shown in Fig. 2a, the monolayer structure consists of two outermost VS$_2$ layers and one Li-ion layer sandwiched in between. To study the dynamic stability of LiV$_2$S$_4$ monolayer, the phonon spectrum has been calculated using the above $D_{3d}$ symmetric structure with perfect in-plane triangular lattice. As shown in Fig. 2b, the existence of imaginary vibration modes indicates dynamical instability of this highly symmetric system. Further modal analysis suggests these high-imaginary-frequency vibrations are mainly contributed by in-plane displacement of V ions, resulting in a moderate deformation of the triangular lattice (see Fig. S2 in supplementary material for more details).

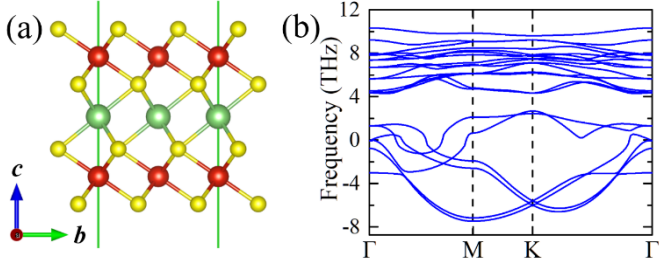

Fig. 2 (a) The side view of the highly symmetric (*P*-3m1) LiV$_2$S$_4$ monolayer. (b) The calculated phonon dispersion curve of LiV$_2$S$_4$ monolayer with ideal triangular lattice.

To find the ground state of LiV$_2$S$_4$ monolayer, both the in-plane lattice constants and the atomic positions have been fully optimized. Supercells are adopted to accommodate the deformation in modelling with the periodic boundary condition. In the optimized monolayer structure, there are three inequivalent V-V bonds, i.e., 3.27 Å, 3.44 Å, and 3.57 Å, as denoted by *s*, *m*, and *l* in Fig. 3a, respectively. The two in-plane V ions with the shortest V-V distance approach each other, presenting typical dimerization behavior, in contrast to the trimerization observed in its bulk counterpart. As expected, these in-plane dimers alternate along (*a*+*b*)-direction and *b*-axis (here, *a* and *b* are the in-plane lattice vectors of the original primitive cell), forming a staggered pattern. Moreover, this in-plane distortion is not as severe as in bulk form. The difference in V-V bond lengths is between 0.13~0.30 Å, which is much smaller than that (around 0.60 Å) of the bulk material.[29,30] The variation in system energy as a function of the degree of dimerization, which has been normalized relative to the ground state distortion, is shown in Fig. 3b. This in-plane distortion can effectively reduce system energy from the undistorted state to the optimized ground state by about 40 meV/V. Besides, the structural symmetry is also reduced from $D_{3d}$ to $C_{2h}$ (*P*2/c space group), while leaving the layered feature untouched.

Fig. 3 (a) Top view of the optimized LiV$_2$S$_4$ monolayer structure. The V dimerization marked by dashed ellipse leads to differentiation of the in-plane V-V spacing. The short (s), medium (m), and long (l) V-V bonds are represented by red, green, and blue lines, respectively. (b) The variation in system energy with the degree of V dimerization, which has been normalized to the optimized intensity of distortion. (c) The calculated phonon spectrum of LiV$_2$S$_4$ monolayer after V dimerization. (d) The MD simulation results for LiV$_2$S$_4$ monolayer. Insets are the top and side views for the structural snapshots of LiV$_2$S$_4$ monolayer before and after simulation at 300 K.

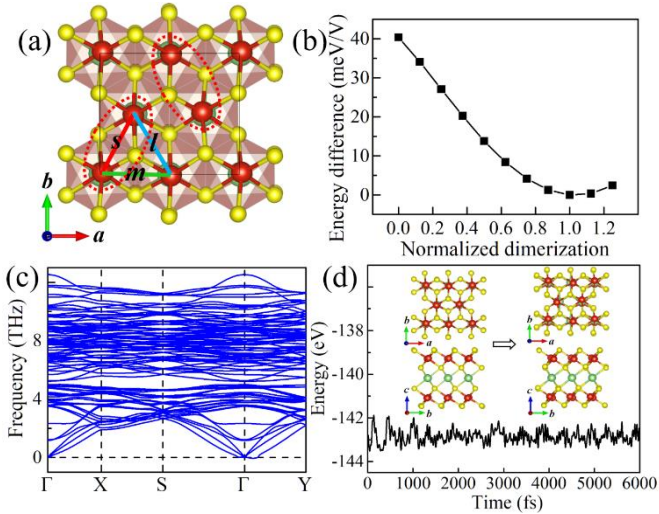

The structure stability of the optimized system has been verified by phonon spectral analysis and molecular dynamics (MD) simulations using the orthorhombic supercell. The calculated phonon spectrum is shown in Fig. 3c, which shows no significant imaginary frequency, indicating the dynamic stability of the structure. Besides, the structure after MD simulations at room temperature (RT = 300 K) shows no severe distortion and the basic crystal framework remains intact despite the existence of thermal fluctuations (see Fig. 3d), further confirming the thermodynamic stability of our optimized structure. It is worth mentioning that the MD simulations also indicate that the in-plane dimerization is not suppressed but rather enhanced. The V-V distance within the dimer is reduced from 3.27 Å at 0 K to less than 3.00 Å at RT. The statistical results of the MD simulations are presented in Fig. S3 of the supplementary material. The mechanism behind this thermal enhancement effect and its potential application is worthy of further exploration.

To investigate the magnetic ground state, four most possible magnetic orders, i.e., ferromagnetic (FM) and two antiferromagnetic (AFM) ones (stripe AFM and zigzag AFM) for the orthorhombic supercell, and an in-plane 120° spin structure (Y-AFM) order for the hexagonal supercell,[40] have been taken into account. The corresponding supercells and magnetic orders are shown in Fig. 4a and Fig. 4b-e, respectively.

Fig. 4 (a) The top view of LiV$_2$S$_4$ monolayer. The unit cell, the hexagonal and orthorhombic supercells are dented with black, blue and green lines, respectively. (b) The 120° spin configuration (Y-AFM) depicted in the hexagonal basis. The in-plane FM (c), stripe AFM (d), and zigzag AFM (e) orders defined in the orthorhombic supercell. The antiparallel spins are represented by dark and light blue arrows, respectively.

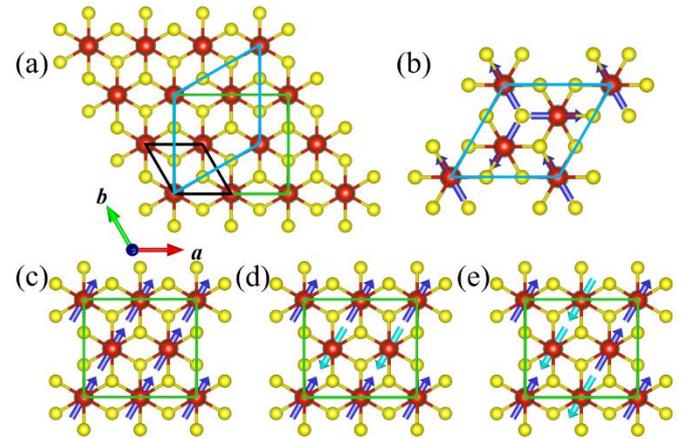

Our calculation results indicate that the FM order is energetically more favorable than the other magnetic orders. The detailed results are summarized in Table S1 of the supplementary material. The magnetic ions in VS$_2$ layer exhibit a significant FM tendency upon peeling from parent bulk to LiV$_2$S$_4$ monolayer. This dimensional reduction induced magnetism can be interpreted on the basis of the $d$ orbital occupation of V. From the change of chemical formula before and after exfoliation, the V ion in LiV$_2$S$_4$ monolayer is in the mixed valence state (V$_2^{7+}$), rather than the original trivalent state in its bulk form. Both the electron hopping among neighboring V with partially filled $d$ orbitals and the $p$ orbital intermediated V-S-V super exchange favor FM coupling and thus give rise to the in-plane FM order.
.

Moreover, since there are two VS$_2$ layers in LiV$_2$S$_4$ (see Fig. 2a), in order to study the interlayer magnetic interaction, the FM layers with AFM interlayer coupling (A-AFM order) has also been taken into consideration. As listed in Table S1, the FM intra- and inter-layer coupling is energetically robust against all the above considered AFM orders and thus is the magnetic ground state of monolayer LiV$_2$S$_4$. The negligible energy difference (0.31 meV/V) between FM and A-AFM order indicates the weak interlayer magnetic coupling, which is consistent with our intuitive expectations since the interlayer V-V spacing is quite large (more than 6.3 Å).

More interestingly, after dimerization, the local magnetic moments show a remarkable divergence within each V dimer, i.e., 1.33 μ$_B$ and 1.88 μ$_B$ corresponding to V$^{4+}$(3$d^1$) and V$^{3+}$(3$d^2$) respectively, indicating the existence of charge disproportionation. What's more, this charge distribution within V dimer synergistically constitutes the in-plane charge ordering. As shown in Fig. 5a, this charge ordering is closely related to vanadium dimerization, presenting a similar distribution pattern which will be discussed in the following paragraph. Moreover, this charge ordering not only lowers the system energy but also opens an energy gap (about 0.65 eV) at the Fermi level (see Fig. 5b and c), resulting in the MIT. Unlike the paramagnetic metal to nonmagnetic insulator transition in bulk LiVS$_2$, this gap opening in LiV$_2$S$_4$ monolayer is strongly spin coupled, while the non-spin polarized DOS exhibits typical metallic behavior (see Fig. S4 in supplementary material). In other words, it is not the in-plane dimerization but rather spin ordering that plays a decisive role in opening the gap.

By comparing the orbital-projected DOS (PDOS) of V ions, the orbital ordering is found to exist in LiV$_2$S$_4$ accompanying the observed charge ordering. As shown in Fig. 5c and depicted in Fig. 5d, the dominant portions of neighboring Vs' 3$d$ orbitals exhibit staggered $d_{xy}$ and $d_{yz}$ ordering. This feature is also demonstrated in Fig. 5a. More precisely, based on our calculation results, the electronic states of the two V ions in one dimer (labelled as V$_1$ and V$_4$ or V$_2$ and V$_3$) are approximately in $d^1$ and $d^2$ configurations, respectively. This finding corresponds to the PDOS, in which the d1 electron occupies the dyz (dxy) orbital of V$_1$ (V$_2$). While for V$_3$ (V$_4$), the $d$yz ($d$xy) orbital is occupied by one $d$ electron, and the other two $t_{2g}$ orbitals are almost equally populated with the other $d$ electron. Therefore, when viewed along c-axis, the local charges on V$_1$ and V$_2$ exhibit a spindle like distribution, while V$_3$ and V$_4$ are more square like. Here, since the $c$-axis does not coincide with the $z$-axis of VO$_6$ octahedron, as illustrated in Fig. 5d, the ELF cross-section does not strictly match the orbital character defined in the $xyz$ coordinate. However, the orderly distribution pattern of local charges is still quite evident. For more intuitive display, the dimers and the V$^{3+}$, V$^{4+}$ ions with staggered distribution are marked by light red dashed ellipses and black dotted lines in Fig. 5a, respectively.

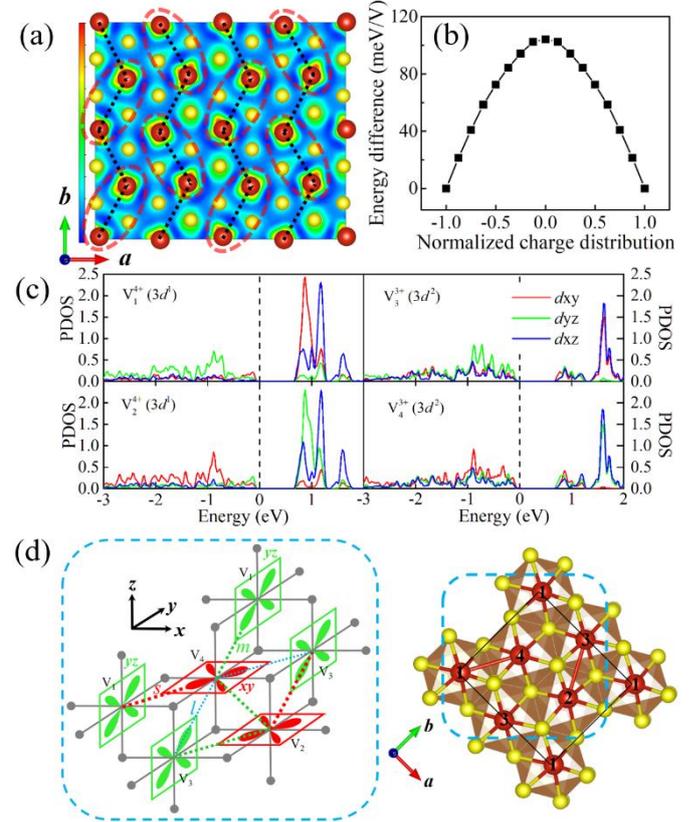

Fig. 5 (a) The 2D cross section of electron localization function (ELF) for LiV$_2$S$_4$ monolayer. (b) The variation in system energy with charge distribution, which is the charge difference between two V ions in the dimer and has been normalized to the optimized value. (c) The orbital projected DOS of V ions. (d) The schematic diagram of orbital ordering on V sites. After dimerization, there are four V ions in LiV$_2$S$_4$ primitive cell, whose labels and nominal chemical valences are marked in (c) and (d).

According to Goodenough-Kanamori-Anderson (GKA) rules, superexchange between two orthogonal $d$ orbitals through intermediate $p$ orbitals (e.g., V$_1$-S-V$_4$, V$_2$-S-V$_3$) leads to an FM interaction. However, the overlap of neighboring $d$ orbitals (e.g., $d_{xy}$ orbitals of V$_2$ and V$_4$) tends to AFM exchange, while the unequal occupation of the other $d$ orbitals (e.g., the $d_{yz}/d_{xz}$ orbitals of V$_2$ and V$_4$) favors FM exchange. All the above factors, together with the effect of unequal V-V distance, result in FM zigzag chains composed of V dimers (e.g., V$_1$-V$_4$-V$_1$ chain, V$_3$-V$_2$-V$_3$ chain, as depicted in Fig.5a) with relatively weak interchain coupling. The strength of FM coupling is strongest between V ions within each dimer, second strongest between adjacent dimers on the same zigzag chain, and then weakest between neighboring chains. The above analysis is in agreement with our in-plane magnetic calculations according to which the zigzag AFM (i.e., intrachain FM order with interchain AFM coupling) is the metastable state, whose energy is only 4.8 meV/V higher than the ground FM state.

In the exfoliated LiV$_2$S$_4$ monolayer, the average valence of V is 3.5. In strongly correlated transition metal systems, analogous to the layered manganite La$_{0.5}$Sr$_{1.5}$MnO$_4$,[41] this non-integer valence induced electronic instability may lead to structural distortion and then be compensated by the distinct occupancies of subsequent ligand states, resulting in the spin, charge, and orbital ordering with lower system energy. While compared to the Mott transition in bulk LiVS$_2$, the MIT in LiV$_2$S$_4$ monolayer is more like a Peierls assisted orbital selective Mott transition proposed in VO$_2$.[42]

## Conclusions

In summary, based on first-principles calculations, the structural, magnetic, and electronic properties of single layer LiV$_2$S$_4$ have been investigated. The in-plane vanadium dimerization has been observed unlike the trimerization found in its parent bulk material. This structural distortion originates from electronic instability aroused by exfoliation along with the strong electron-phonon coupling. Additionally, charge distribution is found in LiV$_2$S$_4$ monolayer with FM coupling within and between vanadium dimers, giving rise to the 2D ferromagnetism. Along with the charge and spin ordering, a gap opens up in the electronic structure, making monolayer LiV$_2$S$_4$ a desirable 2D FM semiconductor, which may play an important role in the ultra-compact spintronic devices. The above novel behaviors predicted in monolayer LiV$_2$S$_4$, including structural dimerization, 2D ferromagnetism, charge and orbital ordering, are expected to be experimentally verified in the near future.

## Author Contributions

The manuscript was written through the contributions of all the authors. All the authors have given approval to the final version of the manuscript.

## Conflicts of interest

There are no conflicts to declare.

## Acknowledgements

This work was supported by the National Natural Science Foundation of China (Grant No. 12274070). Most calculations were done on the National Supercomputing Centre in Jinan and the Big Data Computing Center of Southeast University.